# Nonlinear Transformation of Orbital Angular Momentum through Quasi-phase Matching

Guang-hao Shao (邵光灏), Zi-jian Wu (吴子建), Jin-hui Chen (陈锦辉), Fei Xu (徐飞), and Yan-qing Lu[*] (陆延青)

*National Laboratory of Solid State Microstructures and College of Engineering and Applied Sciences, Nanjing University, Nanjing 210093, China*

We propose and investigate the quasi-phase matched (QPM) nonlinear optical frequency conversion of optical vortices in periodically poled Lithium Niobate (PPLN). Laguerre-Gaussian (LG) modes are used to represent the orbital angular momentum (OAM) states, characterized with the azimuthal and radial indices. Typical three-wave nonlinear interactions among the involved OAM modes are studied with the help of coupling wave equations. Being different from normal QPM process where the energy and quasi-momentum conservations are satisfied, both of the azimuthal and radial indices of the OAM states keep constant in most of the cases. However, abnormal change of the radial index is observed when there is asynchronous nonlinear conversion in different parts of the beams. The QPM nonlinear evolution of fractional OAM states is also discussed showing some interesting properties. In comparison with the traditional birefringent phase matching (BPM), the QPM technique avoids the undesired walk-off effect to reserve high-quality LG modes. We believe the QPM is an efficient way to convert, amplify and switch OAM states in various optical vortex related applications.

PACS numbers: 42.65.Ky, 77.80.Dj, 74.25.Uv

# I. Introduction

In the last decade, optical vortices with orbital angular momentum (OAM) showing helical phase fronts have drawn more and more research attentions [1, 2]. Due to the unique beam profile, optical vortices have been used in optical tweezers, which could trap, move and rotate particles ranging in size from tens of nanometers to tens of micrometers [3, 4]. In addition, optical vortex is also a quite hot topic in information processing and communication areas because data could be encoded into different OAM states. Terabit transmission has been realized based on OAM multiplexing since it supplies a new freedom to carry information [5]. The OAM states even may act as qubits showing many advantages in free space quantum communications [6, 7]. However, for both classical and quantum communication systems, the essential functional units are similar, which should rely on the generation, transmission, transformation and detection of OAM states. So far, various technologies have been used to obtain OAM states with different topological charges, such as spiral phase plate [8], liquid crystal Q-plates [9] and fork gratings [10]. The OAM normally keeps constant during free space propagation thus transmitting OAM states is relatively simple. The detection of OAM is based on the beam intensity and profile measurement with commercial available equipments [11, 12]. However, the transformation of OAM states might be still a challenging work. Inherited from normal communication systems, the OAM involved signal switching, routing, filtering, multiplexing/de-multiplexing and even conversions are quite important for future OAM-loaded networks and processors.

As we know, nonlinear optics is an attractive way to manipulate light beams through light-light interaction. All-optic information processing has been many people's relentless pursuits for years. Investigating the nonlinear optical transformation of OAM states thus exhibits both fundamental research significances and important application values. In 1996 and 1997, M. J. Padgett, L. Allen and their coworkers revealed the conservation law of OAM during frequency conversion. Nonlinear optical crystals such as potassium titanyl phosphate (KTP) were used in their second harmonic generation (SHG) experiments through birefringent phase matching (BPM) [13, 14]. However, BPM normally requires sensitive conditions to satisfy such as angle, wavelength, or temperature. If the beams do not propagate along the crystal's optical axes, the walk-off effect not only affects the conversion efficiency but also may deteriorate the wavefronts. This would be a serious problem for optical vortices due to their unique beam features [15]. On the contrary, quasi-phase matching (QPM) technique could greatly release the limitations of matching conditions through the suitable modulation of nonlinear susceptibility [16, 17]. All involved lights could normally incident and collinearly interact along a crystallographic axis with high efficiency. A typical material is periodically poled Lithium Niobate (PPLN) with all input beams polarized along its z-axis and propagating along the x-axis to use its largest nonlinear coefficient $d_{33}$.

In this paper, the QPM nonlinear optical transformation of OAM is proposed and investigated in PPLN crystals. Besides SHG, sum-frequency generation (SFG) and difference frequency generation (DFG) are also studied. Laguerre-Gaussian (LG) modes are exploited to represent the vortex beams with OAM, which are characterized by the azimuthal index *l* (*i.e.*, the topological charge) and radial index *p* [13, 14]. We find that the QPM is very applicable for optical vortices. The frequency conversion, amplification and switching of OAM modes thus could be realized. However, different from the conventional QPM where only the conservations of energy and momentum are considered, the conservation of OAM also should be taken into account. The total azimuthal indices of the involved lights still keep constant after QPM frequency conversion, while the radial index *p* shows a more complicated evolution. When the conversion efficiency is low in a short PPLN, the radial indices follow the similar rule like that of the azimuthal indices. With higher conversion efficiency in a long PPLN, the pure LG modes of the signal wave may collapse so that the conservation law is violated, providing ways to generate *p* > 0 OAM with *p* = 0 beams. Furthermore, the nonlinear transformation of beams with fractional OAM states is also studied. The related applications are discussed.

## II.     Theoretical analysis

LG modes are characterized by two indices, the azimuthal index *l* and the radial index *p* [13, 14], where *l* is the number of 2π cycles in phase around the circumference and *p* is related to the number of radial nodes. The amplitude $u_p^l(r,\phi,\mathrm{x})$ of a LG mode in cylindrical coordinates is given by

$$u_p^l(r,\phi,\text{x}) \propto \left(\frac{r\sqrt{2}}{\omega(x)}\right)^l \exp\left(-\frac{r^2}{\omega^2(x)}\right)$$
$$\times L_p^l\left(\frac{2r^2}{\omega^2(x)}\right)\exp\left(ik\frac{r^2}{2R(x)}\right) \quad (1)$$
$$\times \exp(il\phi)\exp\left[-i(2p+l+1)\zeta(x)\right],$$

where $\omega(x)$ represents the beam radius at the position of $x$, which is the axial distance from the beam waist. $r$ is the radial distance from the center axis of the beam and $L_p^l$ is the generalized Laguuerre polynomials. $k = 2\pi / \lambda$ means the wave number (in radians per meter). $R(x)$ is the radius of curvature of the beam's wavefronts. $\phi$ represents the azimuthal angle. $(2p + l + 1)\cdot\zeta(x)$ is called Gouy phase, which is an extra contribution to the phase.

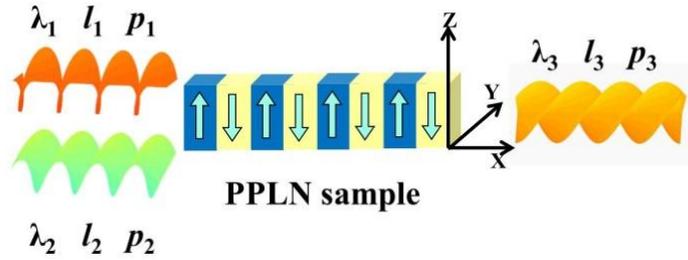

Fig. 1. Schematic diagram representing the QPM frequency conversion of optical vortices by using a PPLN sample.

Figure 1 illustrates the QPM frequency conversion of optical vortices by using a PPLN. To mimic a general three-wave nonlinear interaction process, two optical vortices ($\lambda_1$, $l_1$, $p_1$; $\lambda_2$, $l_2$, $p_2$) are injected and a new optical vortex ($\lambda_3$, $l_3$, $p_3$) is generated as shown in the figure. All these beams are z-polarized and propagate along the x-axis of the sample. To investigate the frequency conversion processes, the coupling wave equations are used to describe the interaction between these waves [18].

$$\begin{cases} \dfrac{dA_{1z}}{dx} = -iK_1 A_{2z}^* A_{3z} e^{-i\Delta kx}, \\ \dfrac{dA_{2z}}{dx} = -iK_2 A_{3z} A_{1z}^* e^{-i\Delta kx}, \\ \dfrac{dA_{3z}}{dx} = -i\cdot ratio \cdot K_3 A_{1z} A_{2z} e^{i\Delta kx}. \end{cases} \quad (2)$$

The subscripts $j = 1, 2, 3$ in Eqs. (2) refer to the involved waves and $z$ represents the polarizations. $A_{jz} = \sqrt{\frac{n_{jz}}{\omega_{jz}}} u_p^l(r,\phi,x)$, where $n$ is the refractive index, $\omega$ is the angular frequencies and $u_p^l(r,\phi,x)$ is described in Eq. (1). $\omega_3 = \omega_1 + \omega_2$ reflects the energy conservation for involved beams. Define $K = \frac{d_{33}g_1}{c}\sqrt{\frac{\omega_{1z}\omega_{2z}\omega_{3z}}{n_{1z}n_{2z}n_{3z}}}$, and $K_{1,2,3}$ are obtained from $K$. $c$ is the light speed in vacuum, $g_1 = 2/\pi$ is a constant and $d_{33} = 25.2$ pm/V is the nonlinear coefficient of Lithium Niobate. For SHG, $K_{1,2,3}$ are all equal to $K$. In addition, the first two equations are identical for SHG, so the *ratio* is 0.5 due to the frequency degeneracy. For SFG and DFG, assume the generated wave is described by subscript $q$, then $K_q$ is equal to the conjugate of $K$ while the other two coefficients keep at $K$. In these cases, *ratio* = 1 as there is no degeneracy. $\Delta k = k_{3z} - k_{2z} - k_{1z} - G_m$, where $k$ represents the wave vector. When QPM is realized, $\Delta k = 0$, which means the momentum mismatch is fully compensated by the reciprocal vector $G_m$ of the PPLN. In another word, the quasi-momentum conservation is satisfied.

As we mainly focus on OAM, we extract the term $\exp[-i(2p+l+1)\zeta(x)]$ from Eq. (1) regarding the evolution of the azimuthal and radial indices. As other terms mainly describe amplitudes, the term $\exp[-i(2p_j+l_j+1)\zeta_j(x)]$ could stand for $A_{jz}$ if only the transformation of OAM is considered. Based on Eqs. (2), for SHG and SFG, the generated term $\exp[-i(2p_3+l_3+1)\zeta_3(x)]$ is obtained as

$$\exp[-i(2p_3+l_3+1)\zeta_3(x)] \\ = \exp[-i(2p_1+l_1+1)\zeta_1(x)] \cdot \exp[-i(2p_2+l_2+1)\zeta_2(x)], \qquad (3)$$

which is just the product of the other two terms. Considering $\zeta_1(x) = \zeta_2(x)$, then $\zeta_3(x)$ should be equal to $\zeta_1(x)$. In this case, both $l_3 = l_1 + l_2$ and $p_3 = p_1 + p_2$ are satisfied, which means the generated light completely takes over the azimuthal and radial indices of the input lights. These two indices are conserved. Similarly, for DFG, the generated $A_{2z}$ is the product of $A_{3z}$ and conjugate of $A_{1z}$. If ignoring terms that are not relevant to the evolution of OAM, we get

$$\begin{aligned}&\exp\left[-i(2p_2+l_2+1)\zeta_2(x)\right]\\&=\exp\left[-i(2p_3+l_3+1)\zeta_3(x)\right]\cdot\exp\left[i(2p_1+l_1+1)\zeta_1(x)\right].\end{aligned} \quad (4)$$

Under the same assumption $\zeta_1(x) = \zeta_3(x)$, the conservation relationship $l_2 = l_3 - l_1$ and $p_2 = p_3 - p_1$ are also obtained.

From the analysis above, we have successfully deduced the conservation law of OAM during three-wave nonlinear interactions using coupling wave equations, which, to our knowledge, has not been well reported in literatures. The nonlinear frequency transformation is based on PPLN utilizing QPM. New phenomenon with high conversion efficiency thus could be expected.

### III. Numerical simulation results

In our analysis above, the conservations of azimuthal and radial indices are expected. However, to obtain the detail information about the frequency conversion process of the optical vortices, such as the intensity and phase distributions, finite element method (FEM) is an effective way. In this case, all terms in Eqs. (1) and (2) should be taken into account for numerical simulation. To be specific, the whole vortex beam area is divided into many small grids. Take the central field of a grid to stand for the whole region then substitute it into the related coupling wave equations. Without loss of generality, SFG is chosen as the first attempt, which means a weak signal vortex and a strong pump vortex incident into a PPLN sample then generate a sum frequency (SF) wave.

Firstly, we set $p = 0$ for the injected signal and pump waves. This is a relative simple case in which the intensity profile just shows a single ring shape. We assume the pump OAM state is z-polarized at 1064 nm with $l = 2$ and $p = 0$. In the meantime, a z-polarized signal vortex is injected collinearly with $l = 1$ and $p = 0$ at the wavelength of 1550 nm. We assume the pump and signal beams have the same waist radius at 100 μm. The corresponding beam divergence angle is about 0.005 radian. The $\zeta_{1,2}(x)$ of the injected waves are approximately equal due to their similar waist radius with a tiny expansion in the sample. Align the sample carefully so that the beam waists are right at the incident front surface ($x = 0$) of the PPLN. The area we considered is 600 μm × 600 μm composing of 200 × 200 grids. Due to the inhomogeneous field distribution, we set the peak intensity of the pump wave at 10 MW/cm$^2$, which is below the damage threshold of Lithium Niobate [19]. The peak intensity of the signal wave is 0.2 MW/cm$^2$. A PPLN with the period of 11.62 μm is selected to satisfy QPM condition at 20 ℃. Assume it has 200 periods so that the whole sample length is 2.32 mm, which is easily achievable for current fabrication technology. In this case, the simulation results are shown in Fig. 2, where the intensity and phase profiles of the involved waves are illustrated.

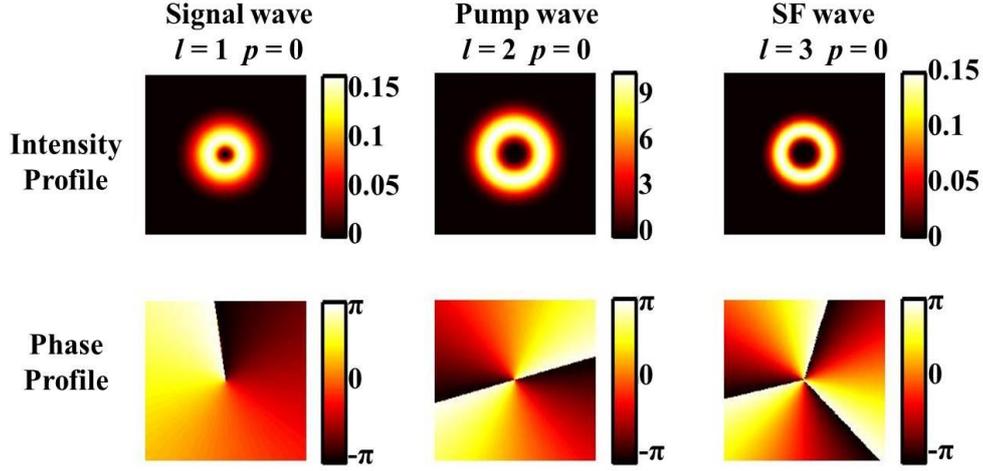

Fig. 2. The intensity and phase profiles of the signal, pump and generated SF optical vortices after a 200-period PPLN sample through QPM.

In Fig. 2, the color bar unit of the field intensity is MW/cm$^2$ and the phase is in radian unit. From the "Intensity Profile" column, there is a central dark hole with very low intensity for each beam. This is the typical pattern of an optical vortex because the phase is hard to define at the center area. Larger dark holes normally correspond to high order of topological charges. It is clear that the SF wave has a relatively larger hole, which means a higher azimuthal index. In addition, in the phase profile, the number of 2π cycles around the circumference is the azimuthal index $l$. We can see that the azimuthal index of SF wave is three. It is just the sum of signal and pump waves' indices, coinciding with the directing reading from the intensity profiles. What's more, all three beams only have single ring in their intensity patterns, meaning their radial indices are all at zero. The conservation laws for both $l$ and $p$ are observed.

Secondly, LG modes with $p > 0$ are also studied. All parameters in our numerical simulation are the same as the previous case except that the radial indices are changed to $p = 1$ for both pump and signal beams. The results are shown in Fig. 3. The color bar units are still MW/cm$^2$ and radian for intensity and phase, respectively.

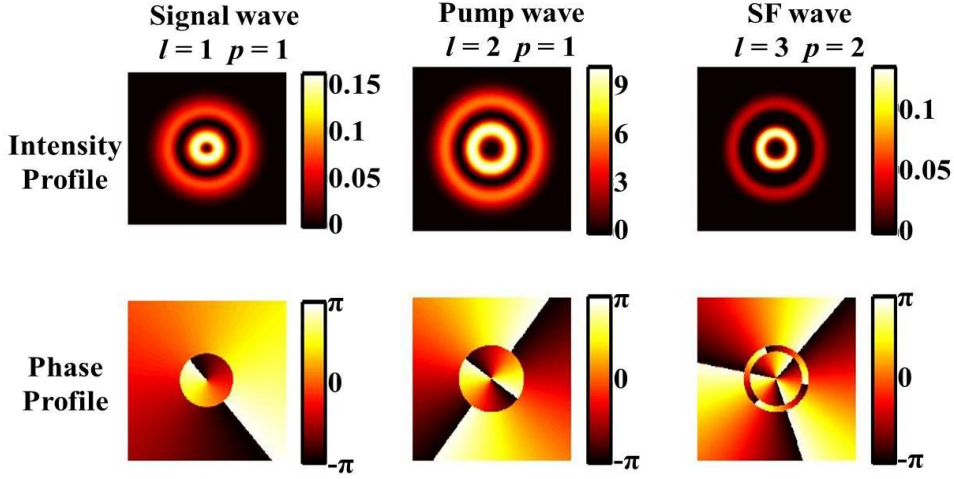

Fig. 3. The intensity and phase profiles of the signal, pump and generated SF optical vortices after passing through a 200-period PPLN sample.

From Fig.3, the intensity profiles of the single and pump waves change remarkably due to the new radial indices, while the central dark holes always exist. There is a clear dark ring between two bright rings in each profile diagram. The radial index thus is read with $p = 1$. For the SF light, $p$ should be 2 if the radial index is conserved. This result is not clear from the intensity profile because one of the bright rings is too dim to be seen. Therefore, the intensity profile is not very accurate to determine the mode indices of a LG mode. However, if we check the phase profile, it is quite evident and convenient. Let's take the pump wave as an example, which is the second one in the second row. From the center to the boarder of the diagram, the phase profile diagram has two zones. In both of the inner and outer zones, the phase undergoes $2 \times 2\pi = 4\pi$, therefore the azimuthal index of the pump beam is 2. And, at the boundary of the inner and outer zones, the phase undergoes a $\pi$-shift, where the corresponding light intensity is zero. Since there are two zones separated by one $\pi$-shift boundary circle in the phase diagram, the radial index of the pump beam is thus 1. Generally speaking, if there are $n$ $\pi$-shift boundary circles from the radial direction, the radial index is just $n$. As a consequence, if we read the phase profile of the SF beam, the radial index $p$ is 2 and the azimuthal index $l$ is 3 as marked in Fig. 3.

From the examples above showing in Figs. 2 and 3, the conservations of both radial and azimuthal indices are testified in this SFG process. This is quite understandable. However, the radial and azimuthal indices are basically determined by the spatial distributions of field intensity and phase. As the nonlinear effects are much more sensitive with respect to the local field, the uneven field profile thus may result in complicated and spatially varied field pattern evolution, which could further influence the involved LG modes and their indices. Some interesting and abnormally phenomena thus are expected, especially when the nonlinear conversion efficiency is high enough. For instance, in previous examples, the peak intensity of signal optical vortex is 0.2 MW/cm$^2$ and the lowest intensity is zero. As light beams propagate in a long PPLN sample employing QPM, the conversion efficiency in some grids could be high while some parts are low, the asynchronous frequency conversion in different grids are established. In some extreme cases, power in some grids of the signal wave may be up-converted to SF wave, while in other grids, lights are flowing back from SF wave to signal wave through a cascading DFG process. In another word, the SFG and DFG processes spatially coupled at the same time in different grids of the beam along a single PPLN sample. Furthermore, the unique phase profile also may play an important role in this complex process superposing of SFG and DFG. It would be interesting to check the transformation of OAM states to see if they still follow Eq. (3) or (4).

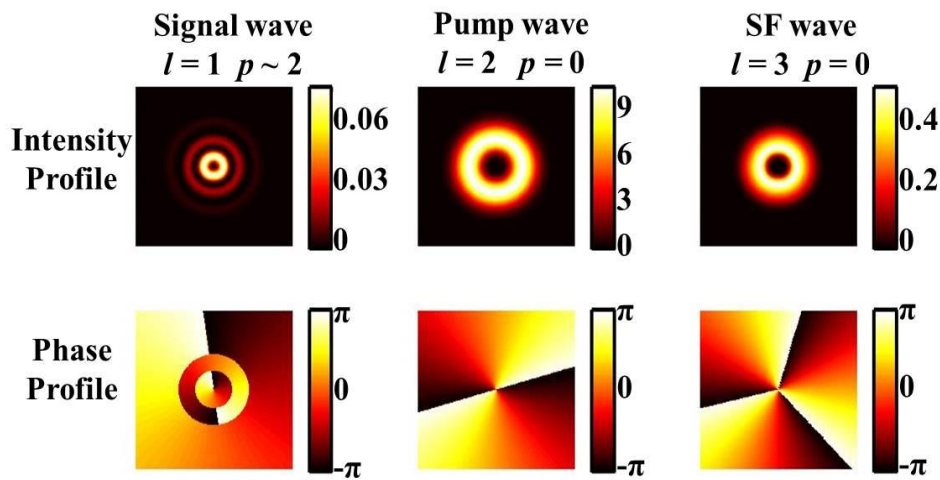

Fig. 4. Output intensity and phase profiles of the signal, pump and generated SF optical vortices after passing through a 600-period PPLN sample.

We still would study a simple case similar to what is shown in Fig. 2. Assume the radial indices $p$ are zero, while the azimuthal indices $l$ are 1 and 2, respectively for the signal and pump waves. All parameters are the same except that a longer sample is used with 600 periods resulting in a sample length of 6.96 mm. The results are shown in Fig. 4. After all lights pass though the sample, a SF vortex ($l = 3$, $p = 0$) is generated along with the original signal and pump wave. We may see that the azimuthal index $l$ of the SF wave is still the sum of pump and signal waves' indices as our expectation. However, the left and regenerated signal wave shows some abnormal transformation of its radial index that does not conserve its original value any more. From both the intensity and phase profiles, the signal wave shows the features of $p = 2$ OAM state. Though it is hard to identify the outermost ring by the naked eyes in intensity profile, we can easily find out that $p = 2$ from phase profile of the signal wave. After propagation through the PPLN, the original LG mode collapse due to the spacially-asynchronous cascading SFG and DFG. In another word, it may supply a new way to generate $p > 0$ beams using $p = 0$ optical vortices.

It should be pointed out that the conservation of azimuthal and radial indices should be obeyed when there is only one frequency conversion process (*e.g.*, SFG) in a PPLN sample. In the case of low conversion efficiency, the power in the signal wave cannot fully convert to the SF wave, let alone the energy flows bi-directionally between injected and generated waves in different parts of the beams. The change of signal's radial index may disappear. High efficiency and collinearly transmission of all beams are thus required to observe the abnormal OAM state transformation. This is just the key advantages of QPM technology in comparison with BPM.

Furthermore, azimuthal indices (*i.e.* topological charges) of OAM states could be integer or fraction [20, 21]. In a more general case, the fractional azimuthal index should be taken into consideration in studying the QPM nonlinear processes. Although a fractional OAM state is not the eigenstate of a vortex beam in the free space, it could be decomposed into the sum of a serial of integer OAM states. The amplitudes of the decomposed integer states are given by Eq. (1). The superposition coefficients $c_{m'}[M(\alpha)]$ is defined as

$$c_{m'}[M(\alpha)] = \exp(-i\mu\alpha) \frac{i \exp[i(l-m')\theta_0]}{2\pi(l-m')} \times \left[ \exp[i(l-m')\alpha](1-\exp(i\mu 2\pi)) \right] \quad . \tag{5}$$

The fractional azimuthal index $l$ is divided into $l = L + \mu$, where $L$ is the integer part and $\mu$ lies between 0 and 1. The orientation of the edge dislocation $\alpha$ is set at zero degree. The angle $\theta_0$ is an arbitrary starting point which defines the interval $\theta_0 \leq \phi < \theta_0 + 2\pi$. Here we set $\theta_0$ at $-\pi$ without loss of generality.

As the fractional OAM state is linear superposition of integer states, it would be interesting to check if it obeys the same transformation rule during frequency conversion. According to our simulation, we found that the fractional azimuthal index conversion also follows the conservation law using the superposition of Eq. (3) or (4). Take $l = 6.5$ as an example, which was obtained by Jörg Götte *et. al.* in 2008 [20]. As the generation of fractional topological charges is complex, we take SHG as an example with fundamental optical vortex at 1550 nm. At 20 °C, a PPLN sample with the period of 18.98 μm is used to satisfy the QPM condition. The whole sample length is 3.80 mm containing 200 periods. Also, the peak intensity is set at 10 MW/cm$^2$. The area we considered is 800 × 800 μm composing of 200 × 200 meshing grids. As the accurate expansion equations are infinite, we only consider the modes whose superposition coefficients $c_{m'}[M(\alpha)]$ is larger than 5 %, meaning $m'$ ranging from 1 to 12. The numerical simulation results are shown in Fig. 5.

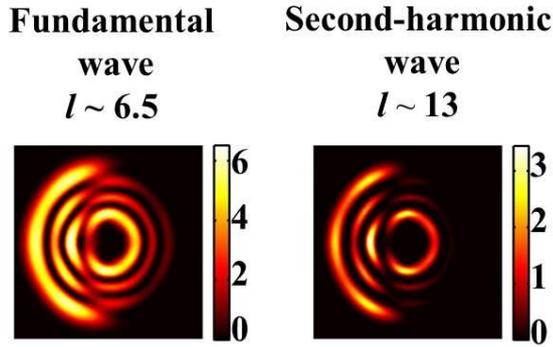

Fig. 5. The intensity profiles of fundamental and second-harmonic optical vortices after passing through a 200-period PPLN sample.

The azimuthal index of OAM could be calculated by [22]

$$\frac{\int u^* \frac{\Delta u}{r\Delta\theta} \cdot r \, dS}{\int u^* u \, dS}, \qquad (6)$$

where $u$ and $r$ have the same meanings as those in Eq. (1). $S$ represents the area of each meshed grid. The units of the color bars in intensity profiles are MW/cm$^2$. Calculating with Eq. (6), the azimuthal indices of fundamental and second-harmonic waves are 6.46 and 12.88, which are close to 6.5 and 13, respectively. The second harmonic light almost doubles the fundamental light. The slight difference is due to the finite expansion. Only if the entire bases are considered, the result may show precise intensity distribution. Therefore, the conservation law of fractional OAM states is still satisfied during the QPM nonlinear conversions.

In this section, four examples are investigated to testify the conservation law of OAM. Both the azimuthal index (*i.e.* topological charge) and radial index have been studied using coupling wave equations through QPM. Besides integer OAM states, the evolution of fractional OAM states have been investigated as well.

## IV. Discussions

For an optical vortex beam with OAM, the wavelength, azimuthal index (*i.e.* topological charge) and radial index are all fundamental parameters. From our analytical deduction and numerical simulation, the QPM frequency conversion is really an effective way to generate and transform the OAM states in a PPLN sample. They could be used to obtain a desired OAM state with new wavelength, topological charge and radial index, while normally governed by the conservation law of OAM. Let's take the SFG as an example. The generated SF and the left-regenerated signal waves are determined by the input signal and pump waves. As long as data are encoded in the signal vortex (or pump beam), they may be transferred to the SF beam at a different wavelength, which is instantly controllable. Our approach thus has promising applications in wavelength routing and switching for vortex-based optical communications. Furthermore, the interesting asynchronous frequency conversion of a signal beam gives some new insights in the QPM frequency conversion processes. In some cases with highly efficient nonlinear frequency conversion, the radial index of the signal wave would not conserve so that we may even obtain $p > 0$ OAM states from a $p = 0$ signal at the same wavelength. This feature may be useful for the multiplexing and demultiplexing of OAM states in OAM-loaded communication or signal processing systems. Even if we move into the quantum world, the QPM transformation of OAM states proposed in this work is very useful. PPLN has already been well-recognized as a quite suitable material to generate entangled photon pairs through spontaneous parametric down conversion. The employment of optical vortex gives a new freedom for photon-entanglement. Both of the azithmual and radial indices may involve. The promising applications of PPLN in this field are very expectable while need extensive studies.

Furthermore, when we talk about second-order nonlinearity, such as SHG, SFG, the related process is very sensitive to the lights' polarization states, which means there is also strict limitation of the optical vortices' polarization states for nonlinear frequency conversion. This is a fundamental feature for BPM nonlinear optics. However, the PPLN or similar materials' domain structure could be designed though agile domain-engineering. In our previous works, we have achieved polarization insensitive SHG, SFG and DFG based on suitable domain designs and assisted with electro-optical modulation [23-25]. Therefore the limitation of involved vortices' polarization states may be released. This is a quite attractive advantage that does not exist in BPM processes. Actually PPLN is such a versatile material with many functions such as nonlinear frequency conversion, electro-optic effect, piezoelectric effect, acoustic-optic effect and so on [26-28]. The coupling and function integration of different effects in a PPLN handling optical vortex may result in some novel phenomena [29], which deserve our future investigations.

## V. Conclusion

In conclusion, we have studied the nonlinear transformation of OAM through QPM in PPLN using coupling wave equations. LG modes are used to represent optical vortex with OAM. Both theoretical analysis and numerical simulation results are given. When a typical three-wave quadratic nonlinear process happens, the conservation law of OAM is revealed, for both azimuthal and radial indices of LG modes. Due to the special intensity and phase profiles of an optical vortex, we also found the spatially asynchronous frequency conversion phenomenon that results in the abnormal radial index change in long PPLN samples with high conversion efficiency. It provides us a new way to generate and modulate an optical vortex's radial index through nonlinear interactions. The extension and applications of our approach are also discussed.

**Acknowledgments**

Guang-hao Shao acknowledges Dr. Jörg Götte for fruitful discussion regarding light beams with fractional OAM. This work is sponsored by 973 programs with contract No. 2011CBA00200 and 2012CB921803, and the National Science Fund for Distinguished Young Scholars with contract No. 61225026.